\documentclass[%
 reprint,
 amsmath,amssymb,
 aip,
 rsi
]{revtex4-2}
\newcommand{\shorteq}{%
  \settowidth{\@tempdima}{-}
  \resizebox{\@tempdima}{\height}{=}%
}
\usepackage{graphicx}
\usepackage{dcolumn}
\usepackage{bm}
\usepackage{tensor}
\usepackage{gensymb}
\usepackage[dvipsnames]{xcolor}
\usepackage{siunitx}
\usepackage{comment}
\usepackage{multirow}
\DeclareSIUnit\mbar{mbar}
\usepackage{float} 
\usepackage{multirow} 

\begin{document}

\bibliographystyle{unsrt}
\preprint{APS/123-QED}

\title{Room temperature buffer gas beam of metastable state titanium atoms}

\author{Jackson Schrott}
    \affiliation{Department of Physics, University of California, Berkeley, CA 94720}
    \affiliation{Challenge Institute for Quantum Computation, University of California, Berkeley, CA 94720}
\author{Scott Eustice}%
    \affiliation{Department of Physics, University of California, Berkeley, CA 94720}
    \affiliation{Challenge Institute for Quantum Computation, University of California, Berkeley, CA 94720}
\author{Dan M.\ Stamper-Kurn}
    \affiliation{Department of Physics, University of California, Berkeley, CA 94720}
    \affiliation{Challenge Institute for Quantum Computation, University of California, Berkeley, CA  94720}
    \affiliation{Materials Science Division, Lawrence Berkeley National Laboratory, Berkeley, CA 94720}

\date{\today}

\begin{abstract}

We produce beams of neutral titanium (Ti) atoms in their metastable $3d^{3}(^4F){4}s$ $a^5F_5$ state by laser ablation into He, N$_2$, and Ar buffer gases. The high temperatures realized in the ablation process populate the $a^5F_5$ level without the need for optical pumping. Remarkably, we observe that Ti atoms in the $a^5F_5$ state survive $1000$'s of collisions with He and Ar buffer gas atoms without being quenched to lower-energy states.  We study the yield of Ti atoms when ablated into buffer gases of varying species and pressure, quantify quenching rates and diffusion cross sections based on simple models, and provide insight into optimal design parameters for an ablation cell. Using a \SI{3.3}{\cm} ablation cell with interchangeable exit apertures, we produce metastable atom beams and quantify their brilliance and velocity distributions as functions of buffer gas pressure.

\end{abstract}

\maketitle

\section{Introduction}

Atomic and molecular beams find numerous applications across the fields of spectroscopy, molecular beam epitaxy, lithography, and ultracold quantum science \cite{scoles_atomic_1988,ramsey_molecular_2005}. Among the techniques developed to produce such beams, the use of buffer gases to entrain or strew a species in a flow of inert gas is notable for its applicability to a wide range of species, including those that evade more conventional approaches like heating in effusive ovens \cite{maxwell_high-flux_2005,stwalley_cooling_2009,hutzler_buffer_2012}.  A buffer gas beam (BGB) operates by introducing a target species of atoms or molecules into a cell containing a pressure of inert buffer gas and extracting some fraction of the species and buffer gas through an orifice. The species can be introduced in a variety of ways including laser ablation\cite{hemmerling_buffer_2014}, electric discharge\cite{babin_characterization_1986}, or gas dosing\cite{wright_cryogenic_2023}.  In the presence of buffer gas, instead of flying ballistically and sticking to the walls of the vacuum chamber, the target species diffuses within and thermalizes collisionally with the buffer gas, allowing for the possibility that species particles make their way to the exit orifice and emerge into a vacuum chamber as a particle beam.

\begin{figure} [t]
    \centering
    \includegraphics[width=\linewidth]{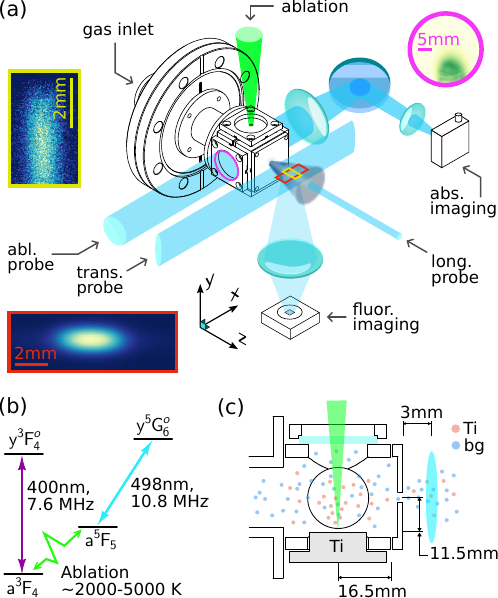}
    \caption{\label{fig:apparatus}(a) In vacuum view of the ablation cell and the various probe beams used in this study. Ablation light enters through the top window of the cell and is focused onto a Ti target at the bottom. Absorption imaging of the plume of Ti atoms resulting from ablation is done on both the $a^3F_4$ and $a^5F_5$ internal states of the Ti atoms. Fluorescence imaging of the atomic beam is performed for atoms in the $a^5F_5$ state. Probe beams are directed transversely and longitudinally through the beam while the scattered light is imaged by a camera below the chamber. (b) Level structure used in this study. The violet \SI{400}{\nm} (teal \SI{498}{\nm}) light addresses the $a^3F_4$ ($a^5F_5$) atoms. The rates indicate the Einstein coefficient ($A_{ki}$) of the transitions in linear frequency units. The broken green \SI{532}{\nm} line indicates that the high temperatures associated with the ablation process significantly populate the metastable $a^5F_5$ state. (c) A closer, cutaway view of the ablation cell. The windows and faceplates of the cell are connected with Viton o-ring seals. The bottom faceplate is made of Ti and protrudes \SI{5}{mm} into the cell. The transverse fluorescence probe is \SI{3}{\mm} downstream of the exit aperture and has a major axis $1/e^2$ beam radius of \SI{2.8}{\mm}.}
\end{figure}

Here, we report on the development of a BGB source for transition metal atoms.  Our motivation in developing this source is to enable laser cooling and trapping of various transition metals (Sc, Y, La, Ti, Zr, V, Nb, Mn, Tc, Fe, and Ru) \cite{eustice_laser_2020}.  This particular application presents two notable challenges:  First, many of the transition metals under consideration are refractory.  Producing atomic beams of such metals by evaporation of an elemental source becomes increasingly challenging as operational temperatures exceed \SI{1800}{\kelvin}  \cite{ross_high_1995}. 
Second, the laser cooling scheme proposed for transition metals typically requires that atoms be prepared in a high-energy metastable state.  The population of atoms produced in this metastable state at sublimation temperatures is small.  While optical pumping can be used in Ti to enrich the metastable state population \cite{schrott_atomic_2024}, this approach adds experimental complexity and must be tailored specifically to each atom and isotope.

We find that a buffer gas beam source, demonstrated here for Ti but likely adaptable to other transition metal atoms, overcomes both these challenges.  Using pulsed-laser ablation of an elemental metal target, we produce rich plumes of atomic vapor within a chamber that is kept at ambient (room) temperature.  We observe the Ti atoms arrive quickly, i.e.\ within 10's of \SI{}{\us},  at a kinetic equilibrium with a room temperature buffer gas of either He, N$_2$ or Ar.  Further, and remarkably, we observe this vapor to be rich with metastable state atoms, not only right after ablation, but also after at least 1000's of collisions with the inert buffer gas in the case of He or Ar. Reduced inelastic collision rates between submerged $d$-shell atoms and filled $s$-shell nobles like He have previously been observed in the ground electronic states of several transition metals, including Ti\cite{hancox_magnetic_2005, hancox_suppression_2005, zygelman_theoretical_2008}, but not in electronically excited states. Allowing the Ti and buffer gas to exit the ablation cell through an orifice, we observe bright atomic beams of metastable Ti, obviating the need for optical pumping.

For convenience, we operate this BGB source at room temperature.  In several previous applications of BGBs, the buffer gas is brought to cryogenic temperatures so that the target species emerges from the buffer gas cell already at very low thermal velocities \cite{hutzler_buffer_2012}.  Such low exit velocities are useful for laser cooling of molecules where the openness of laser cooling transitions allows a molecule to scatter only on the order of $10^4$ photons before the molecule decays into a dark state and no longer scatters laser cooling light \cite{mccarron_laser_2018}.  Hence, the maximum initial velocity from which the molecule can be slowed to zero final velocity is only on the order of 10's of \si[per-mode=repeated-symbol]{\meter\per\second}.  In the present case of Ti and for other laser-coolable transition metals, the laser cooling optical transition has little leakage, allowing very many photons to be scattered before the atom might be pumped into a dark state; for example, a Ti atom is expected to scatter on the order of $10^6$ photons before decaying to a dark state \cite{eustice_optical_2023}.  Therefore, transition metal atoms can be slowed to a standstill even from the velocities characteristic of room temperature (100's of \si[per-mode=repeated-symbol]{\meter\per\second}), and cryogenic buffer gas temperatures are no longer required. While operating at cryogenic temperatures may provide other advantages in terms of atomic extraction efficiency and vacuum quality, in this work we pursue room temperature buffer gas beams due to their significantly reduced experimental complexity.

\begin{figure*} [t]
    \centering
    \includegraphics[width=\linewidth]{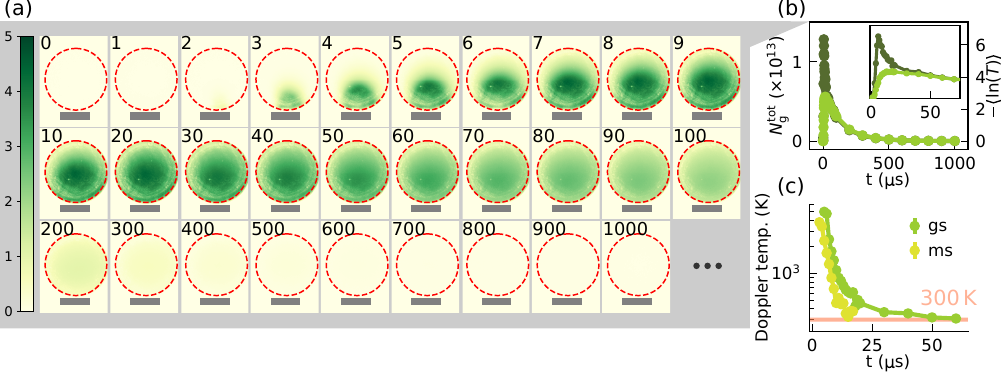}
    \caption{\label{fig:in_cell_imgs}a) Series of absorption images of ground state ($a^3F_4$) Ti atoms ablated into Ar buffer gas at \SI{90}{\micro bar}. The Ti sample is ablated by a single ablation laser pulse, the ablated plume evolves in the buffer gas cell for a variable delay time $t$ before it is imaged with a \SI{1}{\us} duration absorption probe. The atoms are assumed to be unpolarized, and the probe fluence is low so that each atom scatters less than one photon on average.  Optical pumping and saturation effects can be ignored. The images are plotted in log scale so pixel values correspond to local optical depth (OD). The timestamps in the upper left corner of the images are in \SI{}{\us}.  b) 
    By scanning the frequency of the probe laser and recording at each frequency the average OD in the circled region of interest (ROI) of part a, we are able to integrate the OD over the atomic velocity distribution and calculate the total number of ground state atoms in the cell $N_{\mathrm{tot}}$ as a function of time. The total number of ground state atoms within the field of view, $N_{\mathrm{tot}}$, is plotted using the left axis (dark green). On the right axis we plot the average optical depth seen by the probe laser while held at zero detuning (light green). The axes scales are chosen to emphasize that after kinetic thermalization, the optical depth at zero detuning is trivially proportional to $N_{\mathrm{tot}}$ at all later times. We are unable to account for the rapid decrease in $N_{\mathrm{tot}}$ in the first \SI{20}{\micro s}. The inset shows that optical depth at zero detuning becomes an adequate measure of atom number after $\sim$\SI{25}{\us}. c) The effective kinetic temperature is plotted for ground (light green) and metastable state (yellow) Ti atoms. Both metastable and ground state atoms reach kinetic thermal equilibrium with the Ar buffer gas within $\sim$\SI{25}{\us}. }
\end{figure*}

This paper is structured as follows. In Section \ref{sec:apparatus} we describe the experimental apparatus used to study BGBs of Ti, including the construction of our ablation cell. In Section \ref{sec:absorption_imaging} we describe our method for absorption imaging of the evolution of the Ti plume within the room temperature ablation cell. In Sec.\ \ref{sec:yield} we use information from these images to quantify the ablation yield of both ground and metastable state Ti atoms. In Sec.\ \ref{sec:quenching} we use a secondary spectroscopy cell to measure the collisional quenching rate of metastable atoms with the various buffer gases. In Sec.\ \ref{sec:diffusion} we analyze absorption images to determine the cross section for collisions between Ti atoms and buffer gas atoms/molecules through the dynamics of diffusion.  In Section \ref{sec:beam} we characterize Ar BGBs of metastable Ti using different exit orifices and buffer gas pressures. 

\section{Apparatus}
\label{sec:apparatus}
To study BGBs of Ti, we have constructed an apparatus that allows for spectroscopic studies of the dynamics of ablation within a buffer gas cell and also of the properties of extracted beams. Fig.\ \ref{fig:apparatus}a shows the apparatus. An aluminum cell is mounted on the inner (vacuum) face of a conflat flange that has a gas inlet through which we supply buffer gas.  The cell is a \SI{3.3}{\cm} cube with three crossed bores, one of \SI{2.54}{\cm} diameter and two of \SI{1.9}{\cm} diameter. The faces of the cube are machined with grooves to accept \SI{1.5}{\mm} thick viton O-rings. On the front face, an interchangeable exit orifice plate may be connected with an O-ring seal using four screws in the corners of the cube. As described later in Sec.\ \ref{sec:beam}, we use this functionality to study Ti beams from three different types of exit orifice. Interchangeable windows may be mounted on the sides in a similar fashion. The bottom faceplate of the cell is made of Ti and protrudes into the cell \SI{5}{\mm} as shown in Fig.\ \ref{fig:apparatus}c. This plate serves as the ablation target. 

The outer vacuum chamber (not pictured in the figure) is a Kimball Physics DN63 cubic chamber with viewports on the 4 sides adjacent to the ablation cell. On the port opposite the ablation cell, a conical expander and DN160 conflat tee lead to a DN160 Pfeiffer turbo pump (Model No.\ TMU 521P) that pumps the beam region with effective pumping speeds of 190, 90, and \SI{75}{\liter/\s} for room temperature He, N$_2$, and Ar  respectively. The gas flow into the cell is controlled by a MKS Mass Flow Controller (MFC) (Model No.\ GM50A01350085M020) and the pressure in the cell is measured by a thermocouple vacuum gauge near the gas inlet. From knowledge of the molecular conductance between the gauge location and the cell (\SI{4.7}{\liter/\second} for N$_2$), and the gas throughput applied by the MFC, we can relate the pressure measured by our gauge and the pressure in the cell.


We ablate Ti into the buffer gas cell using a Nd:YAG Q-switched laser (Big Sky Model No.\ 532 CLR) that delivers \SI{18}{mJ} of energy at the optical wavelength of \SI{532}{\nm} per \SI{10}{\ns} pulse.  This light is directed through the top window of the buffer gas cell onto the Ti ablation target at \SI{20}{\Hz} repetition rate.  Ablation yields vary significantly based on the freshness of an ablation spot and the inconsistency of the Ti surface. After steering the ablation laser onto a fresh spot on the Ti surface, the ablation yield quickly decreases until reaching a relatively constant level after $\sim$2,000 shots.  After $\sim$100,000 shots the spot is exhausted.  To allow for comparison of ablation yield in the different buffer gases, we ablate a spot for 2000 shots and then take data with all three buffer gases without moving the ablation spot.

We characterize Ti ablation, interactions of atomic Ti with the buffer gas, and the emergence of an atomic beam from the buffer gas source using optical absorption and fluorescence. The population of Ti atoms in the $a^3F_4$ fine-structure level of the ground state term (referenced hereafter as the ground state) is probed with light nearly resonant with the $a ^3F_4 \rightarrow y ^3F_4^o$ transition (see level structure in Fig.\ \ref{fig:apparatus}b).  This \SI{400}{\nm} wavelength light is produced by a external-cavity diode laser (Toptical DLC DL Pro).  Ti atoms in the $a ^5F_5$ state (hereafter referenced as the metastable state) are probed with light nearly resonant with the $a ^5F_5 \rightarrow y ^5G_6^o$ transition, which is the cycling transition that supports laser cooling \cite{eustice_laser_2020}.  This \SI{498}{\nm} wavelength light is produced by frequency doubling the output of a tunable titanium-sapphire laser operating at \SI{996}{\nm} wavelength (Msquared SolsTiS).  Both optical probes can be tuned about the respective atomic resonances by several GHz.  This tunability is used to measure the Doppler broadening of the transitions, and thereby the velocity distribution of atoms in the buffer gas cell and in the atomic beam.

The layout of the laser probe paths and detection optics are depicted in Fig.\ \ref{fig:apparatus}a.  The side windows of the cell allow for large cross section optical beams of \SI{400}{\nm} or \SI{498}{\nm} wavelength to be sent through for absorption imaging of the ablated ground and metastable Ti atoms.

Outside the ablation cell, circularly polarized \SI{498}{\nm} fluorescence probe beams are sent either transversely or longitudinally through the atom beam. The optical density of Ti atoms in the beam was too low to be easily detected by absorption. The magnetic field at the position of the fluorescence probe is estimated to be $\sim$\SI{70}{mG} pointing in the $-y$ direction based on measurements around the vacuum chamber. The fluorescence from atoms was imaged by a camera positioned under the cell as shown in Fig.\ \ref{fig:apparatus}a. The transverse probe beam shape was elliptical with $1/e^2$-intensity beam radii of \SI{0.5}{\mm}  and \SI{2.8}{\mm}.  Given these dimensions are larger than the transverse extension of the atomic beam near the ablation cell orifice, we are assured that the entirety of the atom beam passes through the probe and is detected by fluorescence.

In Sec.\ \ref{sec:quenching}, a secondary spectroscopy cell was used to to measure the collisional quenching rates of metastable atoms with the three buffer gases. To measure these rates, higher stagnant buffer gas pressures and longer diffusion times are required than what we easily achieve in the beam cell. As depicted in Fig.\ \ref{fig:quenching}a, the spectroscopy cell is a NW40 6-way Kwik-Flange cross with a viewport on top for ablation and extended wedged viewports on the sides for probe light. The window extensions prevent the wedged windows from becoming excessively coated in Ti. In Sec.\ \ref{sec:diffusion} (and thereafter) we return to studying the beam cell because its geometry simplifies the interpretation of diffusion dynamics.

\section{Absorption imaging of ablated Ti atomic gases}
\label{sec:absorption_imaging}

Absorption imaging allows us to characterize the early production of atomic Ti by ablation, the thermalization of the momentum distribution of Ti atoms with the buffer gas, and the evolution of the number of Ti atoms in the ground and metastable states.   Fig.\ \ref{fig:in_cell_imgs}a shows a characteristic set of images of ground state Ti atoms ablated into Ar, taken using light tuned to the zero-velocity $a ^3F_4 \rightarrow y ^3F_4$ resonance of $^{48}$Ti (the most abundant isotope).

Our absorption probes are unable to characterize the evolution of the ablation plume within its first few microseconds. Prior studies indicate that ablation produces a hot plasma of Ti, characterized by temperatures as high as 8000 to 16000 K over the first $\sim \SI{500}{\ns}$ \cite{woods_characterization_2014, grojo_plasma_2005}.
The plume is invisible to our probe both because it is mostly composed of Ti ions and because it emerges near the Ti faceplate, outside our field of view. The appearance of an absorption signal at about \SI{3}{\us} marks the first recombination of ions into neutrals \cite{grojo_plasma_2005,sasaki_dynamics_2002}, which takes place already several \si{\mm} from the Ti target.

We determine the momentum distribution of the neutral Ti as a function of time by scanning the laser frequency over the Doppler broadened linewidth of the absorption spectrum.   As shown in Figs.\ \ref{fig:in_cell_imgs}b and c, even at relatively low pressures, the Ti kinetically thermalizes with the buffer gas within $\sim$\SI{25}{\micro s}. With the absorption spectrum now having a constant Doppler linewidth, we can determine the total Ti atom number, in either the ground or the metastable state, just from the optical density observed for an on-resonance probe. We note that the scan over the Doppler linewidth reveals that at very early times a large number of neutrals disappear with a much faster characteristic time than that associated with diffusion (inset of Fig.\ \ref{fig:in_cell_imgs}b). This rapid loss may be caused by the presence of other collision partners in the plume at very early times, including multiply charged Ti ions \cite{grojo_plasma_2005,woods_characterization_2014}. For the remainder of this paper, we ignore this early time behavior and concern ourselves only with the yield of kinetically thermalized atoms measured after \SI{25}{\micro s}. Quantitatively, we obtain the total number of thermalized atoms $N_{\mathrm{g},\mathrm{m}}(t)$, in either the ground (g) or metastable (m) state, by integrating the optical density over the image, dividing by the resonant cross section (for an unpolarized gas, and for the specific transition being probed), and then multiplying by the ratio of the Doppler linewidth to the natural linewidth. 

\section{Ablation yield of ground and metastable states}
\label{sec:yield}

\begin{figure} [b]
    \centering
    \includegraphics[width=\linewidth]{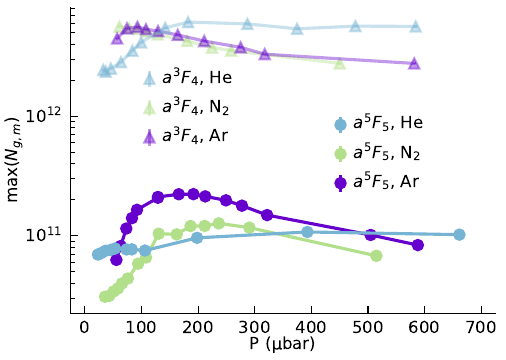}
    \caption{\label{fig:yield}The peak number of kinetically thermalized ground ($a^3F_4$) and metastable ($a^5F_5$) Ti atoms in the cell ($\mathrm{max}(N_\mathrm{g,m})$) detected by absorption probes at various pressures. Between 100\,-\,\SI{500}{\micro bar}, we find the ratio of ground state to metastable state atoms for ablation into He (N$_2$, Ar) buffer gas is on average 37 (23, 15).
    }
\end{figure}

The maximum number of kinetically thermalized atoms $\mathrm{max}(N_{\mathrm{g},\mathrm{m}})$ imaged following ablation into buffer gases of He, N$_2$, and Ar, at a range of buffer gas pressures, are shown in Fig.\ \ref{fig:yield}.  From these data, we discover a striking aspect of the ablation and thermalization process. We find that the ablated Ti atomic gas is remarkably rich in mestastable state atoms.  For Ar buffer gases, we find the metastable atom number to be around 1/15 of the ground state atom number for buffer gas pressures between 100\,-\,\SI{500}{\micro bar}.  For He and N$_2$, the ratios were around 1/37 and 1/23 respectively. 
All these fractions are far higher than one would expect for a Ti gas at thermal equilibrium at room temperature.  At such thermal equilibrium conditions, one would expect $N_\mathrm{m}/N_\mathrm{g} = (g_\mathrm{m}/g_\mathrm{g}) e^{- \Delta E/k_B T} = 4 \times 10^{-14}$, with $g_\mathrm{g} = 9$ and $g_\mathrm{m} = 11$ being the degeneracies of the ground and metastable states, $\Delta E = h c \times 6456 \, \mbox{cm}^{-1}$ being the difference between the metastable and ground state energies, and $T = 300 $ K at room temperature.  Thus, even as the Ti gas equilibrates \emph{kinetically} with the room temperature buffer gas, its internal state distribution does not.  Rather, inverting the Boltzmann relation above, the Ti gas appears to have a much higher effective internal temperature: $T_\mathrm{int} = \frac{\Delta E}{k_B}\left[\ln\left(\frac{g_\mathrm{m}}{g_\mathrm{g}} \frac{N_\mathrm{g}}{N_\mathrm{m}} \right) \right]^{-1}\simeq$ \SI{3200}{\kelvin} for Ar (\SI{2700}{\kelvin} and \SI{2400}{\kelvin} for N$_2$ and He). These temperatures are consistent with the highest temperatures we measure from the Doppler linewidth of the atomic resonance at early times after ablation.  We reason that such high internal state temperatures are inherited from the high initial temperature of the ablation plume. In related works, large populations of metastable Ba atoms were seen during ablation into vacuum, \cite{wang_velocity_2005} and large populations of metastable Nd atoms were observed during ablation into vacuum and noble-gas environments  \cite{rossa_internal_2009}.

\section{Collisional quenching of the Ti metastable state}
\label{sec:quenching}

A second remarkable feature of the ablated Ti atoms is that we find the enriched metastable state population survives in the buffer gas chamber, even as the Ti atoms undergo numerous collisions with He or Ar buffer gas atoms.  By examining the time evolution of the metastable state population, we may constrain the rate at which metastable state atoms relax to lower-energy internal states owing to buffer gas collisions.  We assume the neutralization of the ablation plume is complete within several 10's of \SI{}{\micro s} after the ablation pulse.  From then on, the metastable atom number $N_\mathrm{m}(t)$ may vary owing to three effects.  First, metastable atoms may be depleted from the ablation cell by reaching and sticking to the chamber walls.  Second, the metastable state population may be diminished by exothermic quenching collisions between buffer gas molecules and metastable ($a ^5F_5$) atoms.  Endothermic collisions, by which atoms in lower-energy states can be promoted up into the metastable state, can be neglected given that internal state energy differences are far larger than the kinetic energy $k_B T$ of the buffer gas molecules.  Third, $a ^5F_5$ atoms may be created when Ti atoms in even higher-energy internal states undergo endothermic buffer gas collisions.  We argue that the flux into the $a ^5F_5$ state from such processes is negligible, noting that the $a ^5F_5$ is the highest-energy state within its fine-structure manifold, and that the population in other higher-energy states is likely strongly suppressed by the Boltzmann factor (even at the internal state temperature $T_\mathrm{int}$).  Neglecting this third effect, we expect $N_\mathrm{m}(t)$ to diminish with a rate 
\begin{align}
R_\mathrm{tot}(P) = \zeta_\mathrm{d}/P + \zeta_\mathrm{q}P
\label{eq:loss_model}
\end{align}
where $\zeta_\mathrm{d}/P$ is the loss rate associated with diffusion to the cell wall that is inversely proportional to the buffer gas pressure ($P$), and $\zeta_\mathrm{q}P$ is the loss rate associated with exothermic quenching that is proportional to $P$. 

\begin{figure} [ht]
    \centering
    \includegraphics[width=\linewidth]{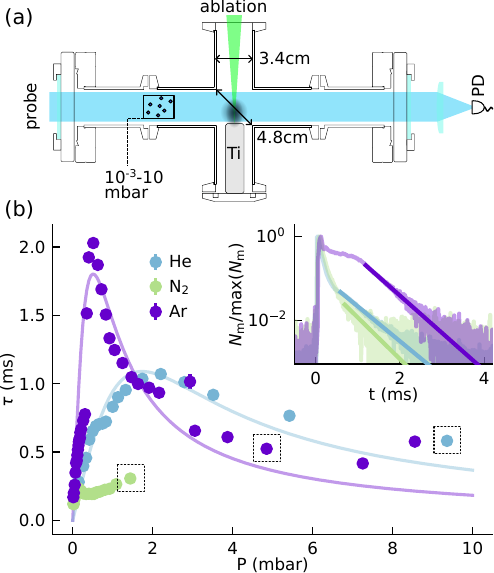}
    \caption{\label{fig:quenching}
    a) In a separate spectroscopy cell of characteristic length \SI{4.8}{\cm}, where the lifetime of atoms is less limited by the diffusion and pump-out times of the cell, long metastable ($a^5F_5$) Ti lifetimes are observed. Buffer gas dosing and pressure measurement are done on the backside of the cell and are not depicted. b) We plot $\tau$, the lifetime of metastable Ti atoms in the spectroscopy cell as a function of the buffer gas pressure $P$ for all three buffer gasses. For He and Ar, we fit the data to Eq.\ \ref{eq:loss_model} in order to extract the collisional quenching rate. In the inset, three peak-normalized absorption traces corresponding to the boxed points in the main figure show the typical signals that we observe. We surmise the Ti atoms survive $\sim3080$ ($\gtrsim$400, 1610) collisions with room temperature He (N$_2$, Ar) atoms.}
\end{figure}

To measure the inelastic quenching rates, we examine ablation of Ti within a separate test chamber, as described in Sec.\ \ref{sec:apparatus}. Over a range of pressures, we fit the exponential decay time ($\tau$) of the metastable atom number at long times after ablation (Fig.\ \ref{fig:quenching}b). By fitting $\tau(P)$ to the model in Eq.\ \ref{eq:loss_model} ($\tau = 1/R_\mathrm{tot}(P)$), we deduce $\zeta_\mathrm{q}$. We measure the following collisional quenching rates of metastable $a^5F_5$ Ti atoms in He and Ar buffer gasses: $\zeta_\mathrm{q}^{\mathrm{Ti-He}}=\SI{240(54)}{s^{-1} mbar^{-1}}$ and $\zeta_\mathrm{q}^{\mathrm{Ti-Ar}}=\SI{540(66)}{s^{-1} mbar^{-1}}$. In N$_2$, the lifetime was not well described by the model and only an upper limit on the quenching rate could be established: $\zeta_\mathrm{q}^{\mathrm{Ti-N}_2}\leq\SI{2260(610)}{s^{-1} mbar^{-1}}$.

Accounting for the buffer gas pressure and the cross section for buffer gas/metastable Ti collisions (derived below), we conclude that metastable Ti atoms survive around $\sim3080$ ($\gtrsim$400, 1610) collisions with room temperature He (N$_2$, Ar) molecules before relaxing to a lower-energy internal state.  Notably, the quenching rate for collisions with N$_2$ may be an order of magnitude larger than with either of the noble gases.

The persistence of high-energy metastable state atoms during collisions with noble elements may be a product of the atomic structure of Ti.  Studies of collisions in the Ti-He system suggest spin relaxing collisions of ground state atoms to lower fine-structure states are significantly suppressed owing to the submerged $d$ shell structure of the Ti ground state ($3d^24s^2$) \cite{lu_fine-structure-changing_2008, zygelman_theoretical_2008, hancox_suppression_2005}.  It may be that metastable state atoms are similarly protected from state-changing collisions by the fact that the outer $4 s$ electron shields the inner $3 d^3$ electrons, preventing the relaxation of the atom both to lower fine-structure states and also to the $3d^2 4s^2$ ground state configuration.


\section{Gas kinetics and measurements of collision cross sections}
\label{sec:diffusion}

We now return to studies of the Ti atoms within the beam cell of Fig.\ \ref{fig:apparatus}a. The evolution of the ablated atoms allows us to determine the cross section for elastic collisions between Ti and room temperature buffer gases.  We employ two methods of analysis, one based on the imaged spatial distribution of the ablated gas, and the second based on measurements of the atom number.

The spatial distribution $n(\mathbf{r}, t)$ of a gas under diffusion evolves according to Fick's 2nd law:
\begin{equation}
    \frac{\partial n}{\partial t} = D \nabla^2 n
\end{equation}
Here, $D$, the diffusion constant, relates to the temperature $T$ and pressure $P$ of the buffer gas by
\begin{equation}
    D = \frac{3 (k_B T)^{3/2}}{16 P \sigma_\mathrm{Ti-bg}} \left( \frac{2 \pi}{\mu} \right)^{1/2}
    \label{eq:diffconstant}
\end{equation}
where $\sigma_\mathrm{Ti-bg}$ is the cross section for collisions between a Ti atom (in the internal state being imaged) and a buffer gas molecule, and $\mu$ is the reduced mass for this collision\cite{massey_free_1934}.

\begin{figure} [t]
    \centering
    \includegraphics[width=\linewidth]{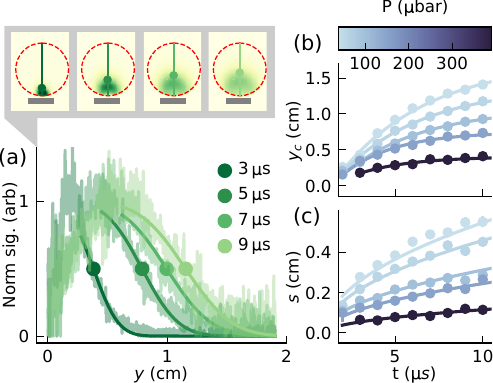}
    \caption{\label{fig:expansion}a) Slices of absorption images at early times (3,5,7,\SI{9}{\us}) are fit with the curve $A(1-\mathrm{Erf}((y-y_c)/s))$. Only points to the right of the maximum of the slice are fit. The callout shows the corresponding absorption images along with a line indicating the 1-d slice and a dot indicating the location of the fit parameter $y_c$. The data shown correspond to metastable Ti ablated into \SI{60}{\micro bar} of Ar. b) Based on fits of ablation plumes like those in a, the time evolution of $y_c$ is plotted for a range of Ar pressures. For each pressure, a linear drag model of the form $y_c = y_f(1-e^{-\beta t})$ is fit to the extracted values of $y_c$ \cite{geohegan_physics_1992}. As pressure is increased, the Ti plume expands a shorter distance into the buffer gas, as shown by the reduced values of $y_c$. c) The time evolution of the ablation plume fit parameter $s$ is plotted for a range of Ar pressures. For each pressure, we fit $s=\sqrt{4Dt'}$, allowing us to extract $D$.}
\end{figure}

We consider two analytical solutions to Fick's law.  First, one may consider the evolution of a sharp one-dimensional front that, at $t = t_0$, separates a uniform density $n_0$ for $y<y_c$ from zero density for $y>y_c$.  Diffusion leads to the following distribution:
\begin{equation}
    n(y,t) = \frac{n_0}{2} \left[ 1 - \mbox{Erf}\left(\frac{y - y_c}{\sqrt{4 D t^\prime}}\right)\right].
    \label{eq:1Dfront}
\end{equation}
where $t^\prime = t - t_0$.  Second, a one-dimensional Gaussian density distribution, centered at $y_0$, evolves as
\begin{equation}
    n(y, t) = n_0 \sqrt{\frac{1}{4 \pi D t^\prime}} \exp\left(- \frac{(y-y_0)^2}{4 D t^\prime}\right),
\end{equation}
where, here, $t_0$ is the extrapolated time where the distribution originates as a delta function.  In either case,  the width of the distribution is proportional to $\sqrt{D t^\prime}$ and can be used to determine $D$.

\begin{figure} [b]
    \centering
    \includegraphics[width=\linewidth]{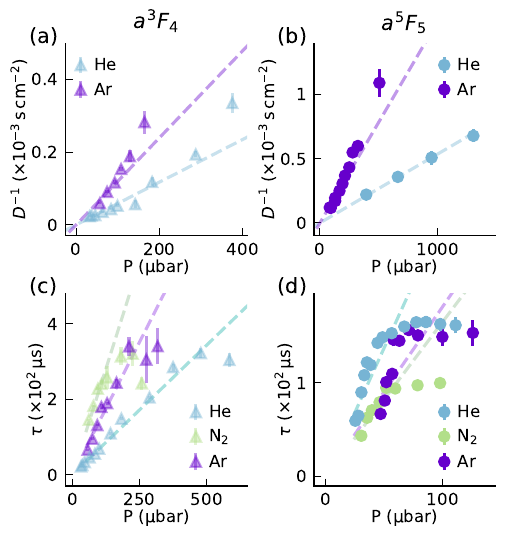}
    \caption{\label{fig:diffusion_fits} a) The inverse of the diffusion coefficients $D$ extracted by fitting the model in Eq.\ \ref{eq:1Dfront} to image slices of the expanding Ti plume. The slope of the linear fits allows for calculation of $\sigma_\mathrm{Ti-bg}$ via Eq.\ \ref{eq:diffconstant}.  b) The late time decay of the Ti population in the cell is compared to the simple diffusion model of Eq.\ \ref{eq:late_diffusion}. The low pressure part of the data is fit to a line and $D$ is extrapolated from the slope per Eq.\ \ref{eqn:tau}. The cross sections gleaned from the fits are shown Tab.\ \ref{tab:cross_sections}.
    }
\end{figure}

We track the spatial distribution of the ablated gas between the time that the gas first propagates into the image ($t \sim$ \SI{3}{\us}, first image in Fig.\ \ref{fig:expansion}a) and when it fills the imaged area ($t \sim$ \SI{9}{\us}, last image in Fig.\ \ref{fig:expansion}a).  Owing to inhomogeneity of the imaged distribution and the limited image field of view, we find it empirically easier to fit an error function to a one-dimensional line cut of the gas.  As shown in Fig.\ \ref{fig:expansion}, we fit error functions of the form $A(1-\mathrm{Erf}((y-y_c)/s))$ with $y_c$ and $s$ treated as free fit parameters at each point in time. The diffusion coefficient is extracted by fitting $s=\sqrt{4Dt'}$. From Eq.\ \ref{eq:1Dfront}, we can then determine $\sigma_\mathrm{Ti-bg}$ from the slope of $D^{-1}$ versus pressure as shown in Fig.\ \ref{fig:diffusion_fits}a. Results of this method for measuring the cross section of both ground and metastable states in He and Ar are shown in Tab.\ \ref{tab:cross_sections}.  We were unable to extract a reliable cross-section for collisions with N$_2$ by this method.

At later times, the ablated gas expands to fill the entire cell, and the total number of ablated atoms diminishes as atoms stick to the cell walls.  To describe the loss of atom number, we consider the solution to Fick's law under the constraint that the density is zero on the surfaces of a square box with equal side length ($L = \SI{3.3}{\cm}$):
\begin{align}
\nonumber n(\mathbf{x},t^\prime) = 
\sum_{l,m,n=1}^{\infty}&A_{lmn}\sin\big(\frac{l\pi x}{L}\big)\sin\big(\frac{m\pi y}{L}\big)\sin\big(\frac{l\pi z}{L}\big)\\
&\times \exp\Big\{-\frac{\pi^2Dt}{L^2}\big(l^2 + m^2 + n^2\big)\Big\},
\label{eq:late_diffusion}
\end{align}
where $(x,y,z) = (0,0,0)$ at a corner of the box, and $A_{lmn}$ is the amplitude of the of the corresponding diffusion mode at $t^\prime=0$. At long times, when the evolution is dominated by the slow decay of the lowest order mode ($l,m,n=1$), the total atom number decays exponentially with time constant
\begin{align}
\tau =  \frac{L^2}{3\pi^2D}.
\label{eqn:tau}
\end{align}

\begin{table} [t]

    \begin{ruledtabular}
    \begin{tabular}{ccccc}
     & $X$ & He & N$_{2}$ & Ar \\
    \hline
    \multirow{2}{*}{Expansion} & $\sigma_{g}$ & 0.94(44) & --- & 0.79(42) \\
    & $\sigma_{m}$ & 0.85(45) & --- & 1.03(49) \\ \hline
    \multirow{2}{*}{Decay time} & $\sigma_{g}$ & 2.98(1.0) & 4.47(1.6) & 2.5(1.0) \\
    & $\sigma_{m}$ & 11.9(4.2) & 3.28(1.1) & 3.24(1.2)\\
    \end{tabular}
    \end{ruledtabular}
    \caption{\label{tab:cross_sections}Diffusion cross sections measured either using the model of Eq.\ \ref{eq:1Dfront} (Expansion) or using Eq.\ \ref{eqn:tau} (Decay time). All cross sections are given in units of $10^{-15}$\si{\square\cm}.}
\end{table}

The measured atom number decay time constants $\tau$ for both ground and metastable state atoms, for different buffer gases and varying buffer gas pressure $P$, are shown in Fig.\ \ref{fig:diffusion_fits}b.  We find $\tau$ increases linearly with $P$ at low gas pressure, consistent with our model.  We use these low-pressure data to determine $D$ and thence cross sections, which are listed in Tab.\ \ref{tab:cross_sections}.  The decay times $\tau$ roll over to lower values at higher buffer gas pressure.  At these high pressures, the ablated gas remains confined near the Ti surface and decays before filling the ablation cell, and the approximation that only the lowest-order diffusion mode is populated at long times is not valid.

In Tab.\ \ref{tab:cross_sections} we compile the diffusion cross sections measured using the models of Eqs.\ \ref{eq:1Dfront} and \ref{eqn:tau}. Across all values, we measure systematically higher cross sections using the decay time method than the expansion dynamics method, perhaps owing to the effective diffusion length $L$ being longer than \SI{3.3}{\cm} on account of the connection between the buffer gas cell and the gas inlet. 
The values obtained for the cross sections are comparable with values previously published for the true ground state of Ti ($a^3F_2$) with Ar (\SI{5.13(30)e-15}{\cm^2}) and N$_2$ (\SI{1.1(5)e-15}{\cm^2}) \cite{ohebsian_kinetic_1980, willmott_reactive_1997}.  

\begin{figure} [h]
    \centering
    \includegraphics[width=\linewidth]{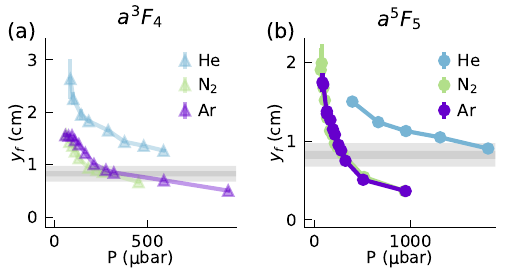}
    \caption{\label{fig:drag_model_fits} a) The stopping distance, $y_f$ of the Ti plume for the different buffer gases. As illustrated in Fig.\ \ref{fig:expansion}b, the position of the plume front, $y_c$, is fit to the function $y_c=y_f(1-e^{-\beta t'})$. The data plotted above are the fit of the parameter $y_f$. The grey stripes show the heights of a \SI{1}{\mm} (dark) or \SI{3}{\mm} (light) aperture used in Sec.\ \ref{sec:beam}. b) Same but for the metastable state.
    }
\end{figure}

In Figs.\ \ref{fig:drag_model_fits} a and b we show the asymptotic distance the Ti plume advances into the cell over a range of buffer gas pressures. The data shown correspond to the fit of $y_f$ in the model $y_c = y_f(1-e^{-\beta t})$ as shown in Fig.\ \ref{fig:expansion}b. The stopping distance of the plume is relevant to the design of ablation cells. As the buffer gas pressure increases, the stopping distance, $y_f$ of the plume decreases. Across all pressures, the stopping distance of Ti in He is larger than in Ar or N$_2$.

In Sec.\ \ref{sec:beam} we replace the Ti target with a taller puck so the target surface sits only \SI{8.25}{\mm} below the exit orifice. This makes the center of the Ti plume more level with the exit orifice at typical operating pressures, ostensibly improving the extraction efficiency into the beam. The new heights of the exit orifices relative to the target surface are highlighted in the figure.

\section{Fluorescence imaging of atom beam}
\label{sec:beam}

We now consider the atomic beam generated from the buffer-gas ablation cell. Using a range of Ar buffer gas pressures and several different exit orifices, we measure the number of metastable atoms in a single pulse of the BGB, $N^{\mathrm{b}}_{\mathrm{m}}$, and the transverse and longitudinal velocity distributions by way of fluorescence spectroscopy. After ablation, atoms extracted into the beam are illuminated by probe light and their fluorescence is collected by a camera underneath the apparatus as shown in Fig. \ref{fig:apparatus}a. We construct velocity distributions by stepping the probe laser frequency to address different velocities of atoms via the Doppler shift.

\begin{figure} [ht]
    \centering
    \includegraphics[width=\linewidth]{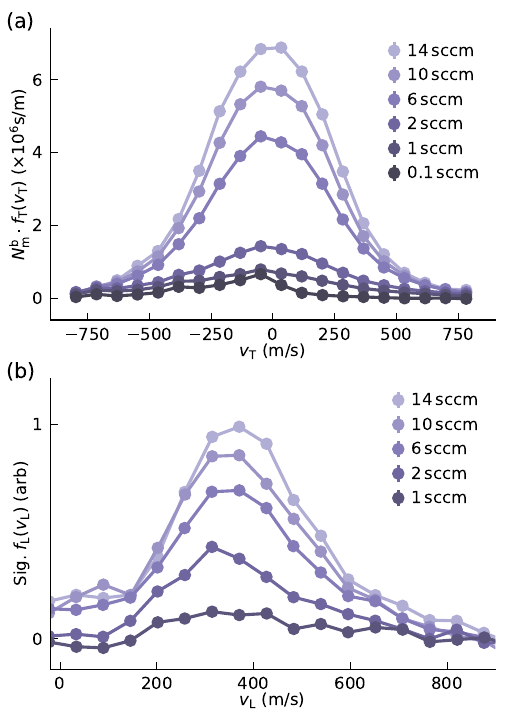}
    \caption{\label{fig:doppler_profiles}a) The transverse velocity distribution of a Ti atom beam emanating from an Ar buffer gas cell with a \SI{3}{\mm} aperture. The distributions are scaled such that the area under each curve corresponds to the number of Ti atoms emitted in the beam after one pulse of the ablation laser. b) The longitudinal velocity distributions of the same atom beam. The relative normalizations correspond to the fluorescence signal size. All the datasets were taken using the same fluorescence beam power. The overall normalization are scaled such that the peak of the signal for the \SI{14}{sccm} data is equal to 1. }
\end{figure}
Fig. \ref{fig:doppler_profiles} shows typical transverse (a) and longitudinal (b) velocity distributions of metastable Ti atoms emanating from an Ar BGB. We observe beams with large fluxes of metastable Ti atoms---up to $N^{\mathrm{b}}_{\mathrm{m}}=\SI{4e9}{atoms/pulse}$---with velocities demonstrative of the ambient motional temperature of the Ti atoms. Importantly, we see large numbers of atoms with velocities within the capture range of a laser-slowing apparatus.

\begin{figure} [h]
    \centering
    \includegraphics[width=\linewidth]{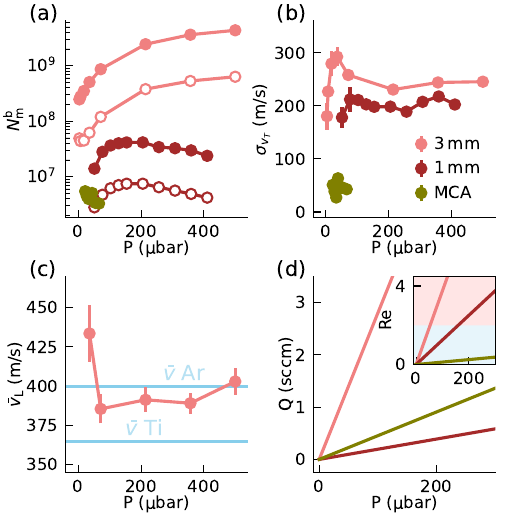}
    \caption{\label{fig:N_per_pulse_and_sigma}Ti atomic beam properties versus ablation cell pressures, $P$. All data is taken with Ar as the buffer gas, and for all data pink, red, and green data refer to the 3 mm, 1 mm and microchannel array (MCA) output apertures respectively. a) Number of metastable Ti atoms emitted into the atomic beam per pulse of the ablation laser $N^{\mathrm{b}}_{\mathrm{m}}$. The unfilled points plot the atom number per pulse that lies within the microchannel array transverse velocity distribution for the larger apertures. b) The transverse velocity width (standard deviation of fitted Gaussian) for the three output apertures. c) The mean forward velocity of the atomic beam emitted by the buffer gas cell with the \SI{3}{mm} diameter aperture installed. Horizontal lines show the average thermal velocity of Ti and Ar at \SI{300}{\kelvin}. d) Buffer gas flow rate $Q$ versus $P$ for the three apertures. An inset shows the approximate Reynolds number in the vicinity of the exit aperture versus $P$ for the different aperture plates. The blue shaded region with $\mathrm{Re}<2$ indicates the regime where effusive flow is expected, while the red shaded region indicates the transitional flow regime where hydrodynamic effects begin to appear. Viscous flow does not occur until Re$\sim$200 \cite{zygelman_theoretical_2008} and does not occur in our system.}
\end{figure}

In Fig. \ref{fig:N_per_pulse_and_sigma} some features of BGBs produced using three different exit orifices are summarized. We use one orifice that is \SI{3}{\mm} in diameter with a wall thickness of \SI{1.5}{\mm} (clear aperture \SI{7.1}{\mm^2}), one that is \SI{1}{\mm} in diameter with a wall thickness of \SI{1.5}{\mm} (clear aperture \SI{0.78}{\mm^2}), and one microchannel array\cite{senaratne_effusive_2015} comprised of 163 holes of diameter \SI{0.2}{\mm} and length \SI{1.5}{\mm} placed on a triangular grid with spacing \SI{0.3}{\mm} (clear aperture \SI{5.1}{\mm^2}).

The choice of exit aperture is an important design consideration as it greatly influences the efficiency with which atoms are extracted into the beam region and also what flow rate is required to achieve a particular operating pressure in the cell. In turn, the flow rate ultimately dictates the gas load on downstream vacuum regions in the apparatus. The total number of atoms in the beam derives from the compounding factors of the ablation yield in the cell, the spatial distribution of Ti atoms in the cell after ablation (i.e.\ the Ti atoms may be stopped too close to the ablation target and diffuse quickly or not be stopped at all), the ratio of the diffusion time to the pump-out time of the chamber, and the possible onset of transitional flow dynamics that tend to increase the extraction efficiency.

The behaviors of the three apertures shown in Fig. \ref{fig:N_per_pulse_and_sigma} highlight some of the trade-offs between these effects. By comparison of Fig. \ref{fig:N_per_pulse_and_sigma}a and Fig. \ref{fig:yield} we see that for the \SI{1}{\mm} aperture, the number of atoms in the beam more or less linearly correlates with the ablation yield, suggesting that the extraction efficiency does not vary significantly over the pressures being explored. This matches our expectation given the diffusion lifetime in the cell is on the order of $100$'s of \SI{}{\micro s} for all the plotted pressures while the pump-out time of the cell is $\tau_\mathrm{pump-out} = V_\mathrm{cell}/C_\mathrm{app}\approx$ \SI{500}{\ms}. Additionally, we expect only minimal transitional flow dynamics to set in for the \SI{1}{\mm} aperture at pressures above \SI{200}{\micro bar} based on the approximate Reynolds number shown in the inset of Fig. \ref{fig:N_per_pulse_and_sigma}d.

On the other hand, the number of metastable atoms emanating from the \SI{3}{\mm} aperture as a function of pressure does not linearly map to ablation yield. Instead more atoms are observed in the beam at high pressures, even as the initial ablation yield starts to decrease. This variation with pressure demonstrates the onset of transitional flow dynamics as predicted by the Reynolds number in Fig. \ref{fig:N_per_pulse_and_sigma}d.

The microchannel array exhibits another operation principle. For a BGB operating in a fully effusive regime ($Re<2$), one optimizes the atom flux by tuning the buffer gas pressure to maximize ablation yield. The microchannel array offers the advantage that one can achieve an exit orifice of clear aperture area $A$ with a vastly smaller molecular conductance than a simple circular aperture, reducing the required flow rate and gas load on the experiment. The conductance of the microchannel array goes as $\propto N d^3/l$ where $N$ is the number of microchannels, $d$ is the diameter of a channel, and $l$ is the length of the channel. With the microchannel array, the transverse velocity profile of the resulting beam is narrowed as $\propto d/l$ so that, in principle, decreasing the exit orifice conductance while maintaining a fixed clear aperture area only comes at the cost of decreasing the number of large transverse velocity atoms in the beam.

Although we do see the expected collimation of the beam by a factor of $0.2/1.5\approx0.13$ as shown in Fig. \ref{fig:N_per_pulse_and_sigma}b, the total atom number from the microchannel array was significantly lower than expected. Given the similar clear apertures of the \SI{3}{\mm} and microchannel array orifices, we expect a comparable number of low transverse velocity atoms in the beams at pressures where the \SI{3}{\mm} aperture is still expected to exhibit molecular flow. Instead we see about 50 times fewer atoms in the microchannel array beam than the \SI{3}{\mm} aperture beam.

For the data described above, the following model was used to infer atom numbers from fluorescence data. The scattering rate $\Gamma$ of an atom with velocity $\mathbf{v}$ is
\begin{align}
    \label{eqn:Gamma}
    \Gamma(\Delta_D) = \frac{\gamma}{2} \frac{s}{1 + s + 4 \Delta_D^2 / \gamma^2}
\end{align}
where $\gamma$ is the FWHM natural linewidth of the transitions, $s=I / I_\mathrm{sat}$ is the saturation parameter of the probe beam, $I_{\mathrm{sat}}$ is the saturation intensity of unpolarized metastable Ti atoms, $\Delta_D = \Delta -\mathbf{k} \cdot \mathbf{v}$ is the effecitve detuning of the light from the atom,  $k=2\pi/\lambda$ is the wavenumber of the probe light and $\Delta$ is the detuning from the zero-velocity probe transition frequency. 
From the scattering rate, the total number of photons scattered during the interaction of a pulse of the atom beam with the transverse probe laser ($N_{\mathrm{emit}}$) was related to the number of atoms in the beam ($N^{\mathrm{b}}_{\mathrm{m}}$) by
\begin{eqnarray}
    N_{\mathrm{emit}} &=&  N^{\mathrm{b}}_{\mathrm{m}}t_{\mathrm{int}}\int\Gamma(\Delta- kv_\mathrm{T})f_\mathrm{T}(v_\mathrm{T})\,dv_\mathrm{T} \nonumber \\
   & \approx & N^{\mathrm{b}}_{\mathrm{m}}t_{\mathrm{int}}\frac{\pi\gamma^2}{4k}\frac{s}{\sqrt{1+s}}f_\mathrm{T}(\Delta/k)
\end{eqnarray}    
where $t_{\mathrm{int}}=2w_{\mathrm{b}}/\bar{v}_\mathrm{L}$ is the average interaction time of atoms with average forward velocities $\bar{v}_\mathrm{L}$ with the probe beam with waist $w_{\mathrm{b}}$. A more complicated analysis\cite{schrott_atomic_2024} that includes effects of optical pumping and the mechanical force of the light on the atoms yields atom-number estimates that differ by less than 20\% with those presented here.  The difference between the methods serves as an estimate of systematic error in our beam-output characterization.

\section{Conclusion}
\label{sec:conclusion}
We have studied the ablation of Ti into room temperature He, N$_2$, and Ar buffer gases, and examined Ar BGBs from three exit orifices. We found that the high temperatures associated with the ablation process produce large populations of metastable atoms. Remarkably, in He and Ar buffer gases, the metastable state population was found to persist through thousands of collisions with buffer gas atoms. The metastable state was quenched more strongly in N$_2$ than in the noble buffer gases.  The characterization of diffusion of Ti within room-temperature buffer gases should aid future experimental design.

This work opens the door to laser cooling and trapping experiments with a number of transition metal elements without the need for optical pumping. We expect the beam source we have presented to be a versatile tool applicable in a wide range of slowing and trapping schemes. 


\section{Acknowledgements}
We give special thanks Jonathan Weinstein and David Lancaster at the University of Nevada, Reno for their help getting us off the ground doing ablation spectroscopy. We thank Diego Novoa for assistance constructing the experimental apparatus and for valuable discussions. We thank Anke St\"oltzel for constructive feedback on the manuscript. We thank Jesse Lopez in the Berkeley Student Machine Shop for his assistance in designing and building the ablation cell. This material is based upon work supported by the Army Research Office (ARO) (Grants No. 049772-001 and 056188-001); the U.S. Department of Energy (DOE), Quantum Systems Accelerator (Grant No. 051120-001); and by the National Science Foundation (NSF) (Grant No. 056175-001).

\section*{Author Declarations}
\subsection*{Conflict of interest}
The authors have no conflicts to disclose.

\subsection*{Author contribution}
\noindent\textbf{Jackson Schrott}: Conceptualization (equal); Data curation (lead); Formal analysis (lead); Investigation (lead); Methodology (lead); Software (lead); Validation (lead); Visualization (lead); Writing – original draft (lead); Writing – review \& editing (equal). \textbf{Scott Eustice}: Conceptualization (equal); Resources (lead); Formal analysis (supporting); Methodology (supporting); Project Administration (lead); Supervision (supporting); Validation (equal); Writing – review \& editing (equal). \textbf{Dan Stamper-Kurn}: Conceptualization (equal); Funding Acquisition (lead); Supervision (lead); Validation (equal); Writing – review \& editing (equal).

\section*{Data availability}
The data that support the findings of this study are available within the article and also from the corresponding authors upon reasonable request.

\clearpage
\bibliography{zot_references2}

\end{document}